# Probing of large interfacial contribution to spin orbit coupling in CoFeB/Ta heterostructure by ultrafast THz emission spectroscopy

**Sandeep Kumar and Sunil Kumar***

Femtosecond Spectroscopy and Nonlinear Photonics Laboratory,
Department of Physics, Indian Institute of Technology Delhi, New Delhi 110016, India.

*Email: kumarsunil@physics.iitd.ac.in

**Ultrafast THz radiation generation from ferromagnetic/nonmagnetic bilayer heterostructure-based spintronic emitters generally exploits the conversion from spin- to charge-current within the nonmagnetic layer and its interface with the ferromagnetic layer. Various possible sub-contributions to the underlying mechanism of inverse spin Hall effect for the THz emission from such structures, need to be exploited for not only investigating the intricacies at the fundamental level in the material properties themselves but also for improving their performance for broadband and high-power THz emission. Here, we report ultrafast THz emission from CoFeB/Ta bilayer at varying sample temperatures in a large range to unravel the role of intrinsic and extrinsic spin to charge conversion processes. In addition to an enhancement in the THz emission, its temperature dependence shows a THz signal polarity reversal if the CoFeB/Ta sample is annealed at an elevated temperature. We extract the behaviour of the spin Hall resistivity, determine the intrinsic spin Hall conductivity contribution in it and compare those with the standard Fe/Pt system. Our results clearly demonstrate a giant interfacial contribution to the overall spin Hall angle arising from the modified interface in the annealed CoFeB/Ta, where a sign reversal in the corresponding spin Hall angle is manifested from the THz amplitude variation with the temperature.**

Spin-orbit coupling (SOC)[1] in solids is the origin of various interesting relativistic phenomena, such as the spin Hall effect (SHE) and its inverse (ISHE),[2, 3] the spin-orbit torque[4] (SOT), Rashba-Edelstein effect (REE) and its inverse (IREE)[5, 6]. Spin-Hall angle is the measurable quantity that determines the charge to spin or spin to charge conversion in the above processes. In spintronic devices, where SHE and its inverse (ISHE) are the leading contributors, the generation, manipulation and detection of the spin current can be effectively done by electrical[7, 8] and optical means[9, 10]. It is always desirous to obtain spintronic materials, their appropriate combinations and heterostructures, which can provide a large spin (charge) to charge (spin) current conversion efficiency so as to implement them in practical applications. For designing such spintronic material structures with large spin Hall angle, it is necessary to understand the underlying fundamental processes. In the spin Hall effect, both the extrinsic and the intrinsic mechanisms play a pivotal role and a distinction between the two can be decisive for specific applications. The intrinsic processes include spin Berry curvature relating to the deflection of spins in the electronic band structure of a perfectly ordered materials[11]. The extrinsic term is related to the disorder-induced scattering events due to the localized impurities. The corresponding contribution in the overall SHE is recognized as skew scattering[12] and side-jump scattering.[13] Disentanglement of each contribution in the total SHE is important so as to enhance the spin-charge (or charge-spin) conversion efficiency and hence the ultrafast terahertz (THz) radiation generation from ferromagnetic/nonmagnetic (FM/NM) bilayer heterostructure-based spintronic THz emitters, where the NM layer is typically a heavy metal. Temperature-dependent experiments for physical property measurements are done routinely for this purpose on the heavy metals.[14-17] By determining the temperature dependence of spin Hall resistivity behaviour (and/or conductivity) and spin Hall angle, the contribution of the dominant mechanism can be quantified. For instance, from spin absorption experiments providing a value of the intrinsic spin Hall conductivity to be $-820 \pm 120$ $(\hbar/e)$ $\Omega^{-1}\text{cm}^{-1}$ in a high-resistive β-phase tantalum (Ta), the intrinsic mechanism is dominant.[14] Similarly, from resistivity measurements on another popular NM heavy metal, platinum (Pt), the dominance of the intrinsic mechanism over the extrinsic one and vice-versa, depending on the spin Hall conductivity, has been determined.[17] In multilayer structures, the interface and its quality can also affect the extrinsic contributions drastically. It has been seen that the extrinsic contribution to SHE and ISHE is many times higher in Py/Pt than that in bulk Pt.[18, 19]

For probing the spin-charge conversion efficiency and the corresponding underlying mechanisms, there exist well-established transport methods,[14, 18, 20, 21] such as the harmonic Hall measurements, ferromagnetic resonance (FMR) based techniques, spin absorption technique, and DC spin Seebeck effect. In recent years, THz emission time-domain spectroscopy has become another more promising technique for these measurements in a non-destructive and contactless manner.[22-24] Either the spin to charge conversion in spintronic structures is directly seen in terms of the THz emission efficiency or THz pulses can be employed in a time-domain spectroscopy to characterize them.[25] For example, temperature-dependent THz emission studies were used to determine the dominant intrinsic contribution to the spin-charge conversion mechanism in Co/Pt bilayer heterostructure.[24] Similarly, it has also been utilized as a probe to identify the dominance of skew scattering at the interface of various spintronic heterostructures.[26] Most recently, laser-induced THz emission spectroscopy has been used as a fingerprint identification experimental scheme to disentangle the ISHE from IREE in Ag/Bi heterostructure,[27] which had otherwise been known before as the Rashba interface material heterostructure having only IREE.[28, 29] Tantalum is a popular choice as the NM layer in many FM/NM based spintronic devices. Even more, Ta is routinely used as a capping or buffer layer. It can be grown in either $\alpha-$ or $\beta-$ or mixed phase, where the spin Hall angle and the resistivity shows a dramatic variation from one phase to the other.[30] A large spin Hall angle value makes this NM heavy metal the popular choice in varieties of SHE and ISHE based spintronic devices. In CoFeB/Ta bilayer heterostructure, a large magnetization reversal[31] and tunnel magnetoresistance[32] have been realised to make this FM/NM combination an appropriate choice in the field of spintronics and also a significant one for the generation of THz radiation by ultrafast optical excitation. Moreover, the annealing of such structure strongly modifies the material and interfacial properties[33, 34] which ultimately affect spin-charge mechanism-related parameters. In addition, CoFeB also owns properties[35, 36] like good spin injection, and very weak sensitivity towards temperature variation for resistivity, magnetization and self-spin Hall effect, etc. Its heterostructure with an NM layer having low bulk spin-orbit coupling strength can, therefore help observe substantial interfacial spin to charge conversion[37] and its temperature dependent behavior.

In this paper, we report temperature-dependent THz emission from thickness optimized CoFeB/Ta and Fe/Pt bilayer spintronic heterostructures, which help us to quantitatively distinguish between distinct contributions to the spin-charge conversion mechanism in CoFeB/Ta spintronic THz emitter. By combining the temperature-dependent resistivity and THz emission measurements, we show that in the as-grown bilayer samples, i.e., CoFeB/Ta and Fe/Pt, the ISHE is driven by the dominating intrinsic contribution. On the other hand, for the annealed CoFeB/Ta sample, a sign reversal of the spin Hall conductivity is manifested from the experimentally observed polarity reversed THz emission below a certain temperature. This peculiar behaviour is attributed to the dominance of the interfacial contribution over the intrinsic bulk one owing to the interfacial modification in the annealed sample.

High-quality thin-film FM/NM bilayer heterostructures of the ferromagnetic and nonmagnetic materials were fabricated by ultra-high vacuum RF/DC magnetron sputtering, and they were used as spintronic THz emitters in a time-domain spectrometer for the results presented here. The thin film deposition order was substrate/FM/NM, and no in-situ substrate heating protocol was used. The CoFeB(3nm)/Ta(3nm) and Fe(3nm)/Pt(3nm) bilayers were deposited on pre-treated high-resistive <100> silicon substrates (HRSi) having thickness Si(380 μm)/$SiO_2$(0.1 μm) in an ultra-high vacuum with a base pressure value lower than $6\times10^{-8}$ Torr. The numbers inside small parentheses represent the thickness of the respective layer in nanometer and for brevity, the same has been removed in the rest of the paper. Under pre-treatment procedure, the substrates were chemically cleaned by double ultrasonication in acetone and isopropyl alcohol for removing various types of impurities from the silicon surface. For the CoFeB/Ta and Fe/Pt bilayers, the thicknesses of FM and NM layers are such that a reasonably high THz signal is obtained from those combinations.[38] A separate set of samples having individual layers of Pt (3nm) and Ta (3nm) were also deposited. Another piece of the CoFeB/Ta heterostructure was post-annealed for 1 hour at a temperature of $350^0$C under the same base pressure value as mentioned above. The crystalline phase, thickness, roughness, etc., were obtained by the optimized growth parameters during the deposition process and also reconfirmed through X-ray diffraction (XRD) and X-ray reflectivity (XRR) measurements.[38] Particularly, the Ta layer was grown in its $\alpha$-phase using the specific optimized growth rate in our experiments.[30, 38] For all the temperature-dependent electrical transport and THz time-domain emission measurements, we have used a closed-cycle helium cryostat system operating in the temperature range of ~5-450 K. The temperature-dependent resistance (R-T) measurements were performed using four-point van der Pauw method. The layout of the home-built THz time-domain spectroscopy setup with the low-temperature cryostat integrated into it, is shown in Figure 1(a). A femtosecond (fs) laser beam of 800 nm central wavelength, 1 kHz pulse repetition rate, and ~50 fs pulse duration is divided into two parts using a 90:10 beam splitter. The strong part is used to optically excite the spintronic emitter mounted on the cold finger of the cryostat chamber. The collimated excitation beam diameter was ~3 mm, while the fluence at the sample point was fixed at ~1 mJ/$cm^2$. At this fluence value, the optical pump induced changes in the reflection and transmission properties of the substrate are nearly unaffected.[39] The emitted THz radiation is then collected by gold-coated $90^0$ off-axis parabolic mirrors and focused onto a 500 μm thick (100)-oriented ZnTe crystal for detection. A high-resistive silicon wafer is placed just after the sample in order to separate the emitted THz radiation from the residual optical beam. The weak portion from the fs laser beam was used as gating beam for the THz detection. These time-delayed (t) gating pulses are made pass through a hole in the last parabolic mirror in the setup to coincide temporally and spatially with the collinearly propagating THz beam onto the ZnTe crystal (Figure 1(a)) for the detection of the latter by electro-optic sampling scheme. More details about the experimental setup can be found elsewhere[40, 41]. The spintronic emitters were magnetized above the saturation magnetic field[38, 42] of the device using an in-situ external magnetic field equipped within the cryostat chamber (Figure 1(a)). The amount of the externally applied magnetic field was kept ~200 mT.

Typical time-domain THz electro-optic signals generated from the as-grown CoFeB/Ta and the annealed CoFeB/Ta bilayer spintronic emitters are presented in Fig. 1(b) for two extreme temperatures, i.e., 300 K and 15 K. Results from experiments on Fe/Pt under the same experimental conditions of optical excitation, detection and these two sample temperatures, are shown in Fig. 1(c). The spintronic heterostructures were optically excited from the NM side of the heterostructure on the substrate, i.e., the order was, light > NM/FM/substrate. The procedure for extracting the THz electric field ($E_{THz}$) from the experimentally measured time-domain electro-optic signal is provided in Supplementary Section S1. Nevertheless, since the gating pulse duration (~50 fs) in our experiments is very small, the electro-optic signal itself can be taken as $E_{THz}$. The THz bandwidth in our experiments on both the Fe/Pt and CoFeB/Ta bilayers is nearly the same, and it is shown in the inset of Fig. 1(c). Various dips in the Fourier spectrum at different frequencies are due to absorption in moisture, as all the measurements reported in this paper have been performed under normal room humidity conditions.[43] For a comparison between the THz generation efficiency of Fe/Pt and CoFeB/Ta, the peak-to-peak amplitude ($E_{pp}$) of the respective $E_{THz}(t)$ signals as presented in Figs. 1(b) and 1(c), respectively, are plotted in the bar plot in Fig. 1(d). For all of our discussion in the following, the $E_{PP}$ has been defined as shown in Fig. 1(c), where $-E_{PP}$ means polarity reversed THz signal and $E_{PP} = 0$ means no THz signal. The THz pulse shape and width were nearly unchanged for all the samples measured in our experiments. Hence, $E_{pp}$ provides an unambiguous measure of the THz amplitude for each case. The detailed temperature-dependent behaviour of the THz emission from Fe/Pt and CoFeB/Ta is discussed later in the paper; however, the major points to highlight from Figs. 1(b-d) are: (i) polarity of the THz signal from annealed CoFeB/Ta ($-E_{PP}$) is reversed to that from the as-grown CoFeB/Ta ($+E_{PP}$), (ii) the THz signal magnitude is temperature-dependent, and (iii) the THz signal magnitude from as-grown or annealed CoFeB/Ta is nearly $1/10^{th}$ of that from the Fe/Pt at both the extreme temperatures of 300 K and 15 K. From Fig. 1(c), we also note that at the room temperature (300 K) the THz signal from the annealed CoFeB/Ta is nearly 1.3 times stronger than that from the as-grown CoFeB/Ta. This particular observation at the room temperature is in close consistency with another study from the recent literature,[34], and hence the enhancement in the THz emission from the annealed CoFeB/Ta sample can be attributed to the recrystallization followed after boron diffusion into the Ta side during annealing.

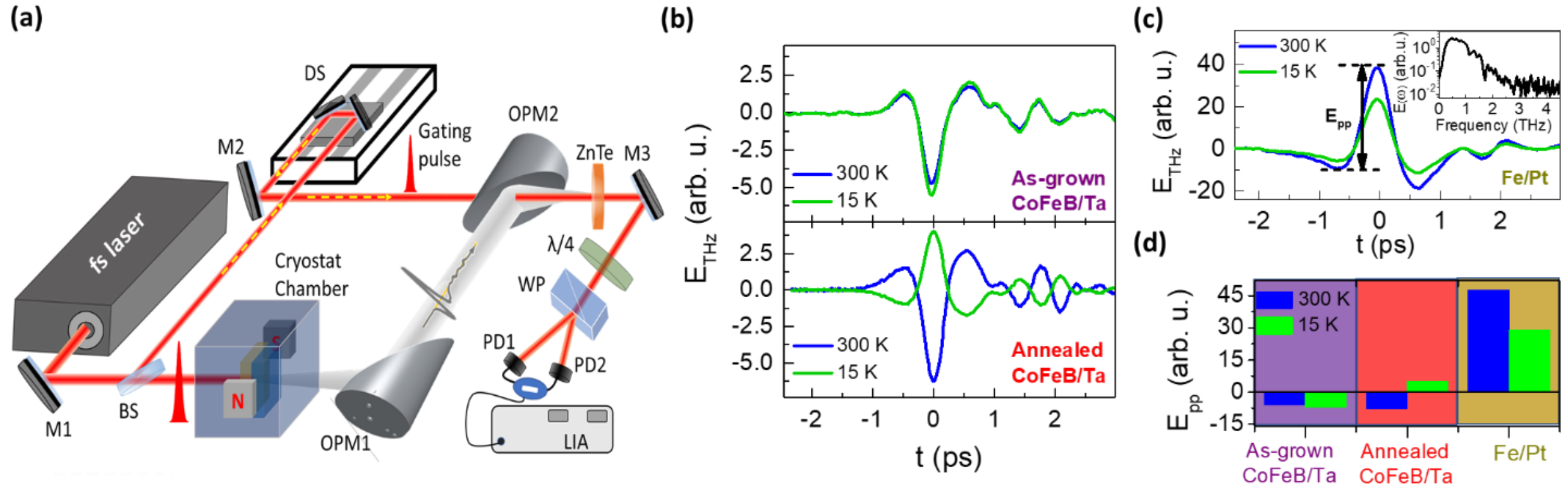

**FIGURE 1.** (a) Cryogenically combined experimental setup for temperature-dependent THz time-domain spectroscopic measurements. (b) Typical THz time-domain signal from as-grown CoFeB/Ta, annealed CoFeB/Ta, and Fe/Pt bilayer spintronic heterostructure at temperature values of 15 K and 300 K. The inset shows the corresponding Fourier spectrum. (c) Peak-to-peak THz signal amplitude ($E_{pp}$) of as-grown CoFeB/Ta, annealed CoFeB/Ta, and Fe/Pt THz emitters for the temperature values 15 K and 300 K. BS: beam splitter, M: mirror, DS: delay stage, OPMs: off-axis parabolic mirrors, λ/4: quarter-wave plate, NC: nonlinear optical crystal, WP: Wollaston prism, BPD: balanced photodiode, LIA: lock-in amplifier.

Figure 2 illustrates the temperature-dependence of the $E_{PP}$(THz) and resistivity (ρ) of the as-grown CoFeB/Ta, annealed CoFeB/Ta, and Fe/Pt heterostructures in a wide range (15-300 K) of the sample temperature. It can be seen from Figs. 2(a,b) that with the decreasing temperature, the THz amplitude for the as-grown CoFeB/Ta ($-E_{PP}$) increases monotonically, while for the annealed CoFeB/Ta, it decreases continuously till a certain temperature only beyond which the polarity gets reversed, and the signal starts to increase again. More precisely, for the annealed CoFeB/Ta, the $-E_{PP}$ signal at high temperatures changes to $+E_{PP}$ signal at low temperatures. On the other hand, the case with annealed Fe/Pt is a separate question in itself and has been addressed partially in a couple of recent articles including ours[44, 45], though the temperature-dependent measurements are yet to be done.

As clear from Fig. 2(c), the THz signal ($+E_{PP}$) simply decreases continuously with the decreasing temperature in Fe/Pt. Such a characteristic temperature dependence in the THz emission has been previously seen Co/Pt also, another Pt-based spintronic THz emitter.[24] The plateau-like feature below ~120 K as seen for the as-grown CoFeB/Ta in Fig. 2(a), is similar to that observed for Co/Pt earlier[24], and such a behaviour can arise from the choice of the FM material layer used. As HR-Si has been shown to have nearly temperature-independent properties[46] e.g., carrier density, relaxation time, etc., in a large range, therefore, we have ignored its impact on the temperature-dependent THz emission behaviour of the spintronic emitters in the current study.

The polarization of the emitted THz radiation from spintronic emitters is basically controlled by the applied magnetic field direction ($\hat{m}$), excitation geometry to define the spin current direction ($J_s$), and the spin Hall angle (θ)[10, 35, 47, 48]. All these parameters are combined in the inverse spin Hall effect mediated charge current relation, i.e., $\vec{J_c} = \theta.(\vec{J_s} \times \hat{m})$. Both, the applied magnetic field and sample geometry were kept the same in our experiments for all the sample types. Therefore, the THz polarity reversal at a certain temperature, as seen in the annealed CoFeB/Ta (Fig. 2(b)), must originate from a similar type of trend in the temperature-dependence, including the sign reversal, if any, of the material's spin Hall angle. The one-to-one correlation between the temperature-dependent behaviour of THz emission from the as-grown CoFeB/Ta and the spin Hall angle of Ta in CoFeB/Ta structure reported in the literature,[16] strengthens the above point. The total spin Hall angle can be expressed as a function of bulk and interfacial part of spin Hall angles, where interfacial spin Hall angle can change from negative to positive or vice-versa, with the temperature variation.[49] This has been further discussed in detail for our case later in the paper.

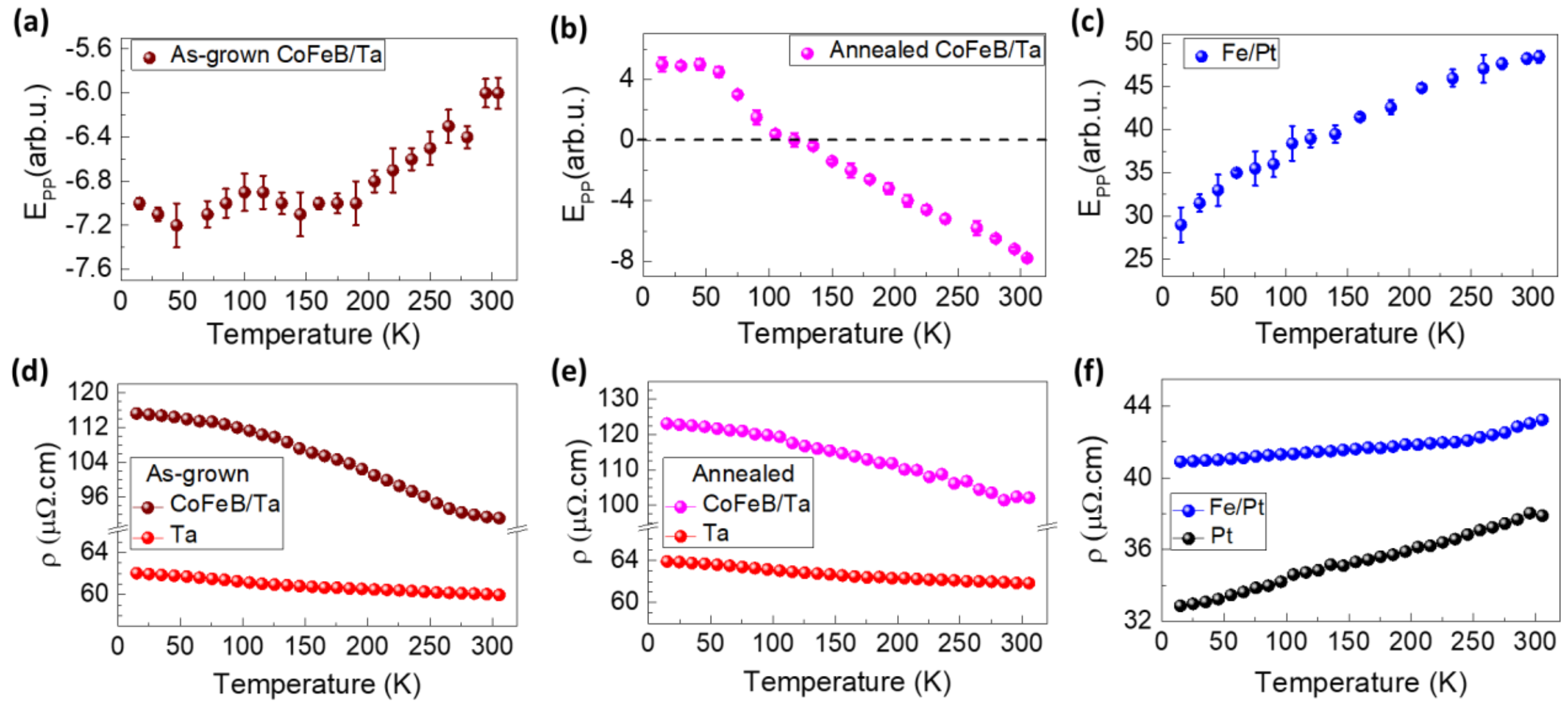

**FIGURE 2.** (a-c) Variation of the magnitude of the THz signal, $E_{pp}$ with respect to the sample temperature varying from the room temperature (300 K) to very low temperatures for the as-grown CoFeB/Ta, annealed CoFeB/Ta, and Fe/Pt spintronic THz emitter, respectively. (d-f) Electrical resistivities (ρ) of the three bilayer samples and only the NM layers in them, all measured using the four-point van der Pauw method in the entire temperature range.

For the NM heavy metal (NM or HM = Ta and Pt in our case), the spin Hall angle is related with the spin Hall resistivity ($\rho_{SH}$) and the electrical resistivity ($\rho_{NM}$) via the relation, [14, 17] $\rho_{SH} = (\theta \, . \, \rho_{NM})$. The temperature dependence of both the $\rho_{SH}$ and $\rho_{NM}$ can help in determining the behavior of θ in the entire temperature range for a given material structure. [17, 50] Figures 2(d-f) show the experimentally measured temperature dependence of the resistivity (ρ) for all the three samples under study, i.e., as-grown CoFeB/Ta, annealed CoFeB/Ta, and Fe/Pt bilayers as well as the respective NM layers (Ta and Pt) by using four-point van der Pauw method. The same could not be done for the CoFeB and Fe alone because of the rapid oxidation problem with these FM layers. However, it is possible to determine the resistivity of these two FM layers from the resistivity of the corresponding bilayers and the counterpart NM layers in them by using the parallel resistor model for the thin film bilayer heterostructures.[51] According to this model, the effective resistivity of the bilayer is given by

$$\rho_T = (d_1 + d_2)\frac{\rho_1\rho_2}{d_1\rho_2 + d_2\rho_1} \tag{1}$$

In the above, $d_{1,2}$ and $\rho_{1,2}$ represent the thickness and the resistivity of the individual layers in the bilayer heterostructure. Thus, an estimated mean value of the resistivity for Fe at room temperature is ~50 μΩ.cm. This value for CoFeB in the as-grown CoFeB/Ta or annealed-CoFeB/Ta is determined to be ~162 μΩ.cm, which is well-matched with the literature.[16]

The longitudinal resistivities of the Pt layer and Fe/Pt bilayer, as seen in Fig. 2(f), decrease with the decreasing temperature in a way that is typical for metallic films.[17, 24]. On the other hand, for Ta layer and CoFeB bilayers in Figs. 2(d-e), the resistivity increases linearly with the decreasing temperature. The negative temperature coefficient of resistance for Ta layers in Figs. 2(d-e) matches well with the behaviour observed in the α-phase Ta below a critical layer thickness of 4 nm[50] and also in β-phase Ta[16], and it can be related to thermally activated charge transfer processes. Amongst all the phases, the α-phase Ta is the low-resistive phase. [16, 30] The mean value of the resistivity being just ~60 μΩ.cm (Figs. 2(d-e)) together with the results from the XRD measurements (see Supplementary Section S3), we confirm that the Ta layers in our case are grown in the α-phase. The resistivity of the FM layers (either Fe or CoFeB, in the present case) is nearly temperature independent throughout the temperature range considered here.[16] Therefore, the temperature dependence in the resistivities of the bilayers, is expected to arise from that of the Pt in Fe/Pt and the Ta in CoFeB/Ta. However, a small difference in the temperature-dependent behaviour of the resistivity of the bilayers as compared with the Pt or the Ta layers alone, is observed in Figs. 2(d-e), and hence, it must originate from the nature of the interface. [52] In fact, we can also note a minute difference between the temperature-dependent resistivities of the as-grown CoFeB/Ta and annealed CoFeB/Ta in Figs. 2(d,e). As, the CoFeB layer alone does not have any temperature dependence[16], such differences are related to the interface modifications in the systems.[33]

Now, we present the results from the THz emission spectroscopy to show that the above-mentioned difference in the resistivity of the as-grown CoFeB/Ta and annealed CoFeB/Ta is indeed related to interface related mechanisms. Only in a few recent studies, temperature-dependent THz time-domain spectroscopy has been employed as a probing tool to investigate the possible microscopic origin of the spin to charge conversion mechanism in the spintronics heterostructures.[24, 26] The amplitude of the generated THz field from the FM/NM bilayer is related to the spin current density and the spin Hall resistivity via a relation[24] as $E_{THz}(\omega) = \rho_{SH} \cdot \left(\frac{1}{d} \cdot \frac{\rho_{FM/HM}}{\rho_{HM}}\right) \cdot J_s(\omega)$. The THz conductivity of the metallic spintronic thin films changes negligibly in a large frequency range of ~0-4 THz[53]. Moreover, because the metallic layers with nanometer thicknesses have small parallel conductivity, the ratio $\frac{\rho_{FM/HM}}{\rho_{HM}}$ remains close to unity in the entire frequency range. Therefore, the magnitude of the emitted THz field and the spin current density can be directly compared in both the time and the frequency domains, and the equation becomes

$$E_{THz}(t) = \rho_{SH} \cdot \left(\frac{1}{d} \cdot \frac{\rho_{FM/HM}}{\rho_{HM}}\right) \cdot J_s(t) \tag{2}$$

Like before, here, $\rho_{SH}$ and $\rho_{NM}$ represent the spin Hall resistivity and longitudinal resistivity, respectively, of the NM layer. Similarly, $\rho_{FM/NM}$ represents the longitudinal resistivity of the FM/NM bilayer having the NM layer on the top from whose side the optical excitation takes place, *d* is the thickness of the bilayer. This relation connecting the dynamic variables (electric field and spin current) and static properties (resistivities) is valid only under the quasi-static approximation.[24] From Eq. 2, it is evident that the THz electric field amplitude majorly depends on the spin Hall resistivity and spin current density, and hence any temperature-dependence in the magnitude of the $E_{THz}$ can be directly related to that of the $\rho_{SH}$ and the $J_S$, provided the ratio between the electrical resistivities of the FM/NM bilayer and the NM layer, $\left(\frac{\rho_{FM/NM}}{\rho_{NM}}\right)$ remains constant. Indeed, this is true in the present case, i.e., $\left(\frac{\rho_{CoFeB/Ta}}{\rho_{Ta}}\right)$ or $\left(\frac{\rho_{Fe/Pt}}{\rho_{Pt}}\right)$ are nearly constant in the entire temperature range in Figs. 2(d-f), a behavior seen even in similar other types of FM/NM bilayers in the literature.[24] Although the change is just a few percent, but still we have included this factor in our calculations later on. Moreover, the spin current in Eq. (2) is also mostly temperature-independent because of two main reasons: (i) the magnetic phase transition temperature (Curie point) of the FM layer is much above than the experimental temperatures used,[54] and (ii) the characteristic parameters relating to the magnetization dynamics to govern the spin current are invariant with respect to the temperature.[24] Therefore, the temperature-dependency in the magnitude of the THz signal can be directly attributed to the temperature-dependent variation of the spin Hall resistivity.

As pointed out earlier, not only the intrinsic but also the extrinsic mechanisms can contribute to the overall behavior of the spin Hall resistivity in a given FM/NM spintronic heterostructure. By including both the intrinsic and the extrinsic contributions, in general, the temperature-dependent spin Hall resistivity can be expressed[14, 17, 55] as

$$\rho_{SH}(T) \ = \ \sigma_{HM}^{int.} \cdot \rho_{HM}^2(T) + \alpha_{ss} \cdot \rho_{0,HM} + \sigma_{SJ} \cdot \rho_{0,HM}^2 \tag{3}$$

Here, the first term on the right-hand side is the weightage of intrinsic contribution relating to the intrinsic spin Hall conductivity, $\sigma_{NM}^{int}$ and the longitudinal electrical resistivity, $\rho_{NM}$ of the NM heavy metal layer. The second and third terms represent the extrinsic contributions: the second term is for skew scattering and the third term is for side-jump scattering, both of which have been pictorially described in Fig. 5. The skew scattering term is characterized by skew scattering angle, $\alpha_{ss}$ and the residual resistivity, $\rho_{0,NM}$ of the NM heavy metal layer, while, the contribution from side-jump scattering is via the corresponding side-jump spin Hall conductivity parameter, $\sigma_{SJ}$. Equations (2) and (3) can be combined for a specific time at which the THz signal is maximum, for example, to analyze the temperature-dependent contributions from the intrinsic and the extrinsic mechanisms to the generation of the THz signal for each of the FM/NM bilayers under study. The resultant can be expressed as

$$\rho_{SH} \propto E_{PP} \cdot \left(\frac{\rho_{HM}}{\rho_{FM/HM}}\right) = \sigma_{HM}^{int.} \cdot \rho_{HM}^2 + \sigma_{SJ} \cdot \rho_{0,HM}^2 + \alpha_{ss} \cdot \rho_{0,HM} \tag{4}$$

From the experimentally known values, $E_{PP} \cdot \left(\frac{\rho_{HM}}{\rho_{FM/HM}}\right)$ *vs* $\rho_{HM}^2$ results are presented in Figs. 3, 4 and 5 for the three bilayer heterostructures, Fe/Pt, as-grown CoFeB/Ta, and annealed CoFeB/Ta, respectively. In each case, a linear fit using Eq. (4) has been used to estimate various intrinsic and extrinsic scattering parameters as desired. The slope of the linear fit provides, $\sigma_{HM}^{int}$, while the intercept provides the value of $\sigma_{SJ} \cdot \rho_{0,HM}^2 + \alpha_{ss} \cdot \rho_{0,HM}$. It can be seen from Figs. 3(a), 4(a) and 5(a) that the linear fits using Eq. (4) are quite reasonable for all three bilayer samples, and hence the intrinsic contribution is confirmed. However, different values of the intercepts amount for the difference in the extrinsic contributions in each one of them for the spin to charge conversion mechanism, which is now discussed in detail in the following paragraphs.

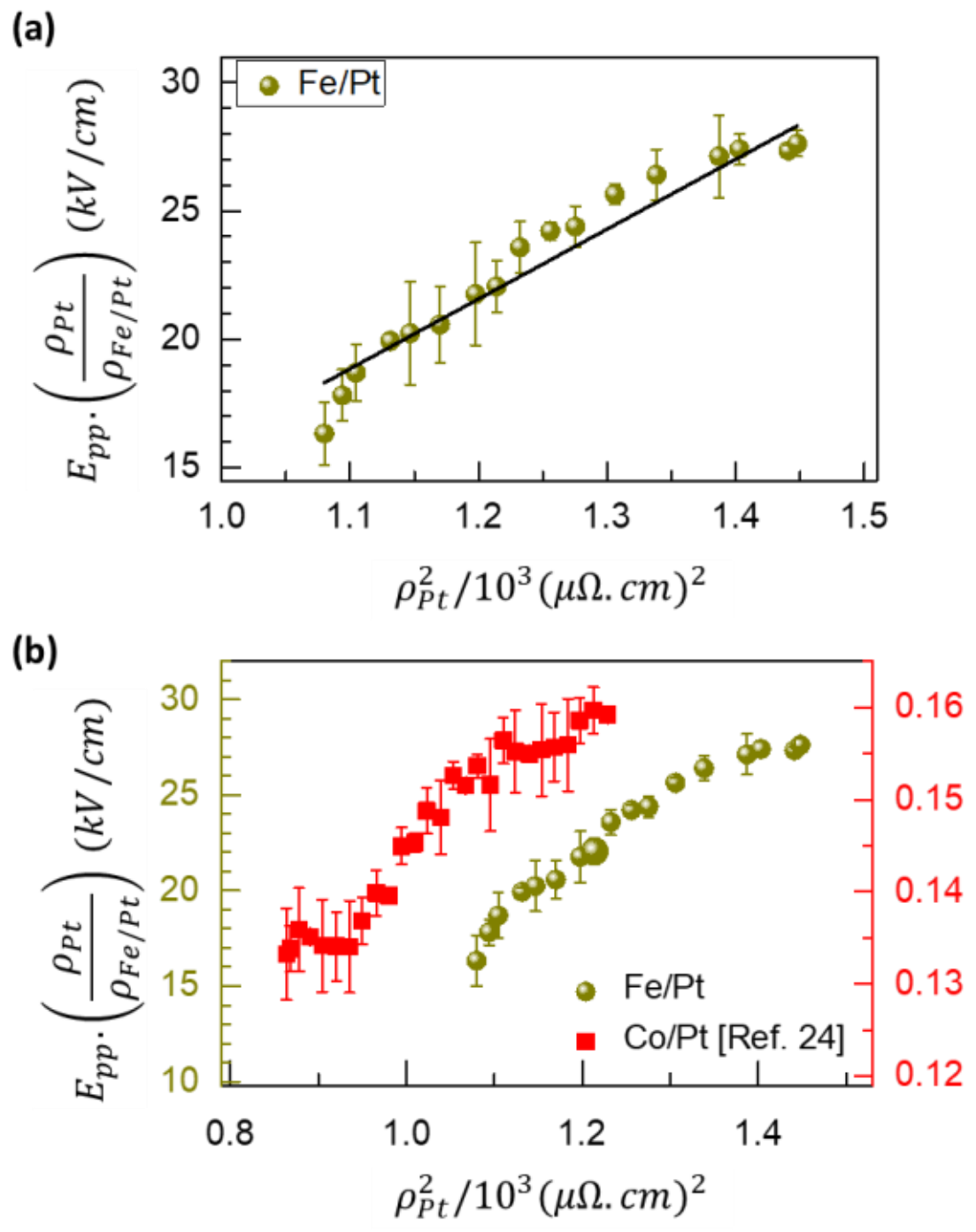

**FIGURE 3.** (a) The spin Hall resistivity of Pt, $\rho_{SH}^{Pt}$ from the experimentally measured THz amplitude, $E_{PP}$ and the squared electrical resistivities $\rho_{Pt}^2$ for the Fe/Pt bilayer. The solid black line is a linear fit to the experimental data using Eq. (4). The values of slope and intercept are 27.2 and -11.1, respectively. (b) Comparison of the temperature-dependent behaviour of the data for Fe/Pt with that for Co/Pt, the latter taken from ref. [24] confirming the dominating intrinsic spin-charge conversion mechanism in the Pt-based FM/NM bilayer spintronic THz emitters.

The temperature-dependent behaviour of the spin Hall resistivity, $\rho_{SH}^{Pt}$ for Pt that is proportional to $E_{PP}.\left(\frac{\rho_{Pt}}{\rho_{Fe/Pt}}\right)$ for the Fe/Pt, has been presented in Fig. 3(a), where the solid line is fit to the $E_{PP}.\left(\frac{\rho_{Pt}}{\rho_{Fe/Pt}}\right)$ vs $\rho_{Pt}^2$ using Eq. (4). From the fitting, we estimate the value of the intrinsic spin Hall conductivity of Pt to be $\sigma_{Pt}^{int} \sim \left(\frac{\hbar}{e}\right) 413\ \Omega^{-1}\text{cm}^{-1}$. Please see Supplementary Section S2 for the details of the calculations, where a constant value of the spin current,[10] was used. The value of the spin Hall conductivity of Pt estimated in the current study is slightly smaller than the range reported in the literature.[17, 56] It may be noted that for the determination of the exact value of the intrinsic spin Hall conductivity, one requires a predetermined value of the spin current density, $J_s$ also, which itself can weakly depend on the temperature.[57] However, the linearly increasing behaviour of the spin Hall resistivity of Pt in Fe/Pt resembles with that of the other Pt-based bilayer FM/NM structures and hence the dominant role of the intrinsic spin-charge conversion mechanism in them is confirmed.[24] For comparison, a result for Co/Pt from the literature[24] is plotted in Fig. 3(b), and the resemblance between the two can be clearly seen. A small and constant horizontal offset in the entire range between the data for the two Pt-based FM/NM bilayers, i.e., Fe/Pt (current study) and Co/Pt from ref.[[24]] arises from the difference in the material combinations and their resistivities. The intrinsic origin of the spin-charge conversion in Pt, as confirmed from our study, is also consistent with the other theoretical[15] and experimental[17, 18] studies on similar Pt-based FM/NM bilayers.

Except for the THz signal polarity reversal, the behaviour of the $E_{PP}.\left(\frac{\rho_{Ta}}{\rho_{CoFeB/Ta}}\right)$ vs $\rho_{Ta}^2$ data for the as-grown CoFeB/Ta, as shown in Fig. 4(a), appears to be similar to that for Fe/Pt. Like before, a linear fit as shown by a solid line in Fig. 4(a) using Eq. (4), provides the experimental value of the intrinsic spin Hall conductivity for the α-phase Ta layer, which comes out to be $\sigma_{Ta}^{int} \sim \left(\frac{\hbar}{e}\right)(-46)\ \Omega^{-1}\text{cm}^{-1}$. This value is in close consistency with the literature,[58, 59] where the spin Hall conductivity for α-phase Ta has been reported to be in the range of ~ –(50-250) $\Omega^{-1}\text{cm}^{-1}$. It is noteworthy to note that the value for the β-phase Ta[14] is relatively much higher (~ –800 $\Omega^{-1}\text{cm}^{-1}$ ), i.e., nearly one order of magnitude higher than the α-phase Ta. Obviously, this difference in the $\sigma_{Ta}^{int.}$ values is due to the difference between the resistivities in the two phases of Ta. In Fig. 4(b), we have re-plotted the $E_{PP}.\left(\frac{\rho_{Ta}}{\rho_{CoFeB/Ta}}\right)$ vs $\rho_{Ta}^2$ data for the α-phase Ta (current study) to compare it with $\rho_{SH}^{Ta}$ vs $\rho_{Ta}^2$ for the β-phase Ta, the latter taken from ref. [14]. Like before, a constant horizontal shift between the data for the CoFeB/Ta(α) and CoFeB/Ta(β) is because of the difference in the resistivities of the Ta layers in the two phases, particularly, the one-order smaller resistivity of Ta(α) than Ta(β). Nevertheless, nearly one to one similarity in the trends of the two plots is quite evident from Fig. 4(b), thereby confirming the predominance of the intrinsic origin for the spin to charge conversion mechanism in the as-grown CoFeB/Ta system.

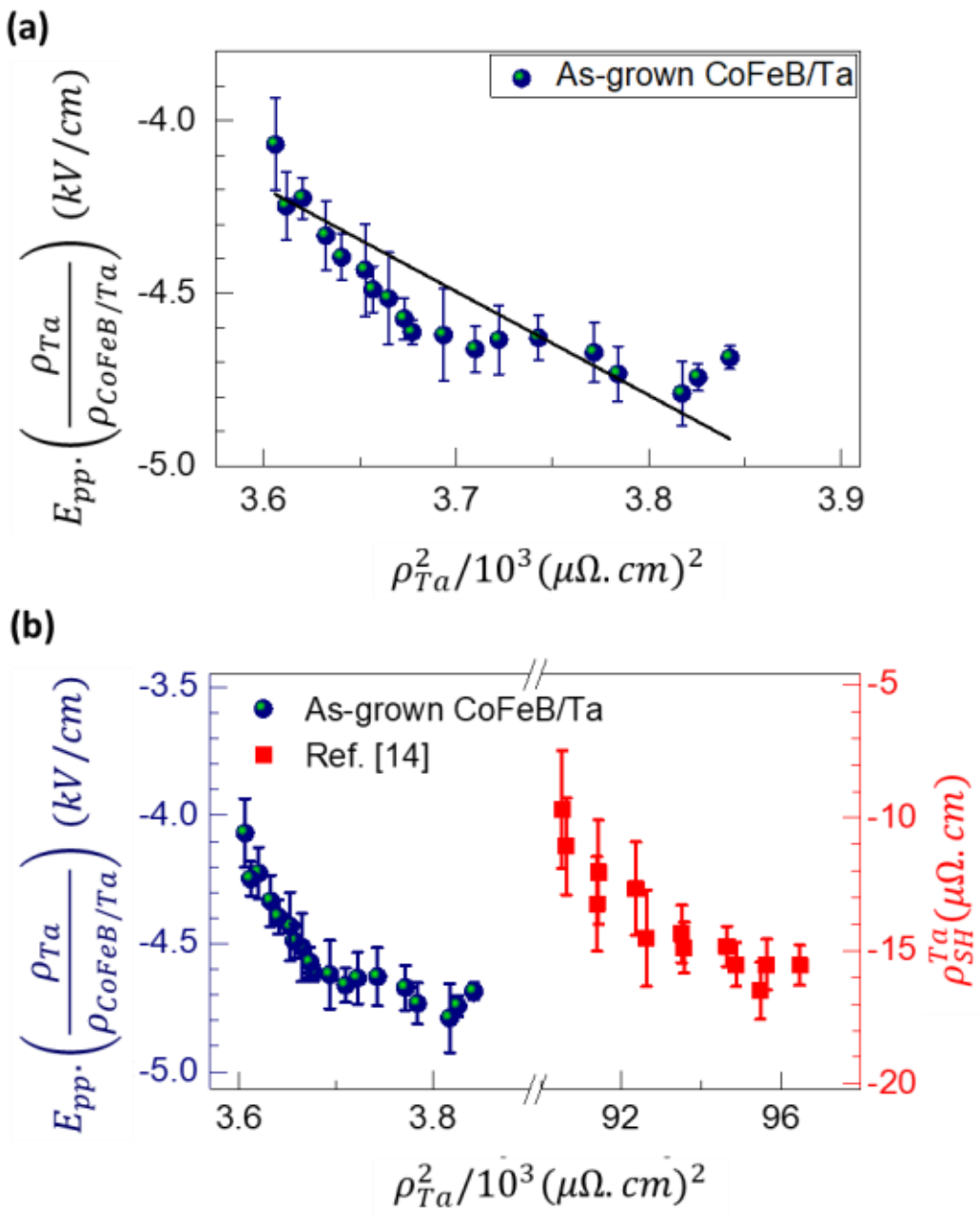

**FIGURE 4.** Spin Hall resistivity, $\rho_{SH}^{Ta} \propto E_{PP}.\left(\frac{\rho_{Ta}}{\rho_{CoFeB/Ta}}\right)$ of the α-phase Ta in the as-grown CoFeB/Ta spintronic THz emitter. (a) $E_{PP}.\left(\frac{\rho_{Ta}}{\rho_{CoFeB/Ta}}\right)$ as a function of $\rho_{Ta}^2$. The solid black line represents fit to the data using Eq. 4. The values of slope and intercept are -3 and 6.6, respectively. (b) Comparison between the spin Hall resistivity of a-phase Ta from the current study against that from ref. [14].

In the case of the annealed CoFeB/Ta sample, the variation of the spin Hall resistivity, $\rho_{SH}^{Ta} \propto E_{PP}.\left(\frac{\rho_{Ta}}{\rho_{CoFeB/Ta}}\right)$, as shown in Fig. 5(a), still follows a linear relation with $\rho_{Ta}^2$. The zero-line crossing of the graph arises from the THz polarity reversal below a certain temperature as shown in Fig. 2(b). From the slope of the linear fit, value of the intrinsic spin Hall conductivity, $\sigma_{Ta}^{int} \sim \left(\frac{\hbar}{e}\right)(377)\ \Omega^{-1}\mathrm{cm}^{-1}$ is obtained. However, the intercept of the linear fit on the $\rho_{Ta}^2$-axis is much different, larger by at least one order than that for the Fe/Pt as well as the as-grown CoFeB/Ta. Since the intercept of the linear fit is a measure of $\sigma_{SJ}.\rho_{0,HM}^2 + \alpha_{ss}.\rho_{0,HM}$ through Eq. (4), therefore, the results suggest that the extrinsic contributions to the overall spin-charge conversion in the annealed CoFeB/Ta is quite significant. Both the extrinsic and intrinsic spin-orbit coupling mechanisms are schematically shown in Fig. 5(b) for duly reference. The extrinsic effect includes the side-jump and skew scattering mechanisms, which are related to the scattering events from impurity/defect centres in the material or its interface with another material.[12, 13]

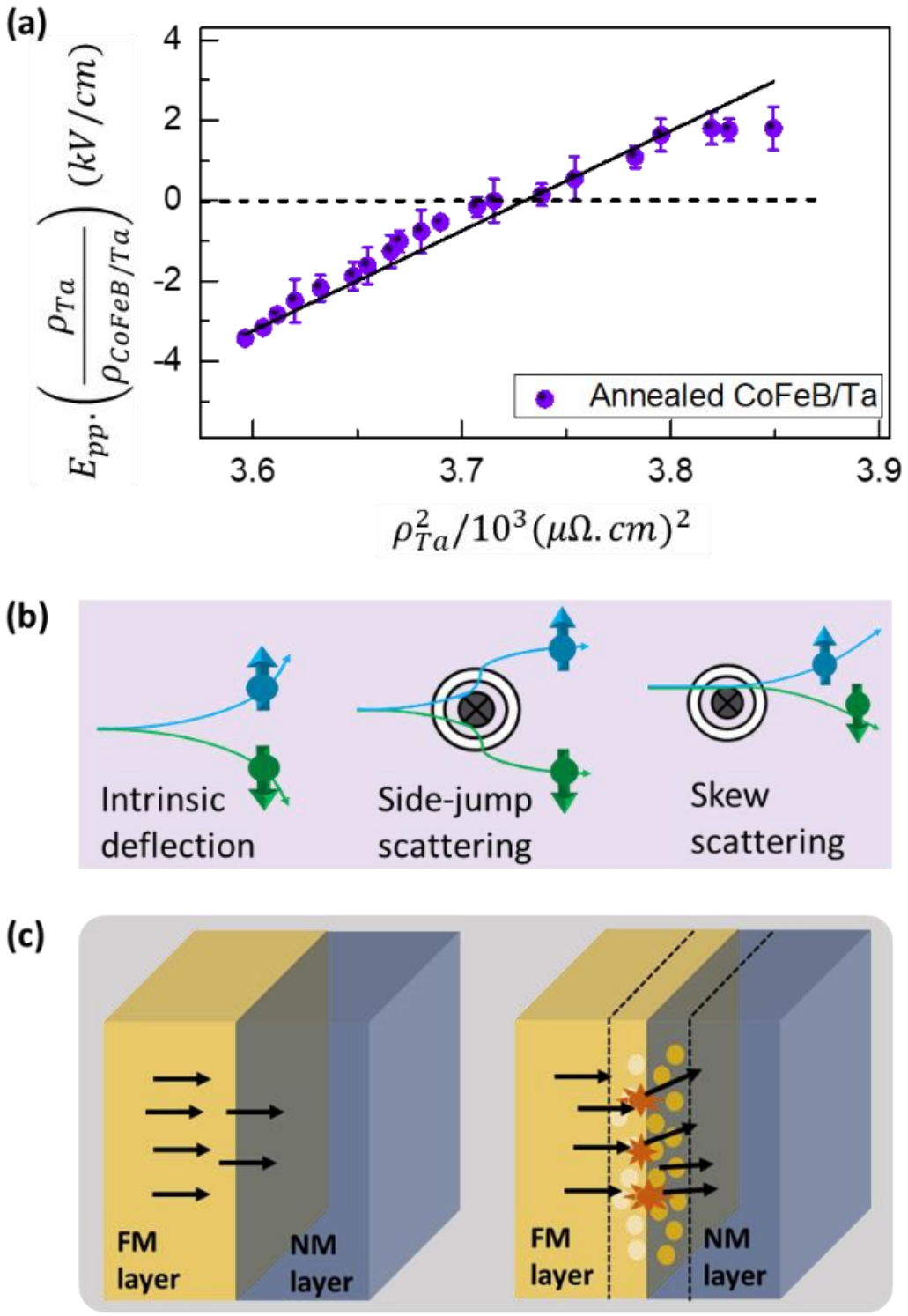

**FIGURE 5.** (a) $E_{PP}.\left(\frac{\rho_{Ta}}{\rho_{CoFeB/Ta}}\right)$ as a function of $\rho_{Ta}^2$ in the annealed CoFeB/Ta bilayer structure. A black solid line represents a linear fit to the data using Eq. 4. The values of the slope and intercept are 24.9 and -92.3, respectively. (b) Schematics for the deflection of the opposite spins in a heavy metal contributing to the spin-charge conversion. The deflection takes place due to either the intrinsic spin-orbit coupling mechanism or by extrinsic mechanisms relating to side-jump and skew scattering from impurities/defects. In a given material, both the intrinsic and the extrinsic mechanisms can contribute to the overall spin-charge conversion. (c) Schematic illustration of the annealing effect in the CoFeB/Ta bilayer in which B diffusion across the interface can alter the interfacial properties and hence the THz emission significantly.

In the inverse spin Hall effect for the THz emission through the usual relation, $\vec{J_c} = \theta.(\vec{J_s} \times \hat{m})$ , a sign change in the THz polarity below a certain temperature, in the case of annealed CoFeB/Ta as shown in Fig. 2(b), implies that the spin Hall angle or the spin Hall resistivity, $\rho_{SH}^{Ta} = \theta^{Ta}.\rho_{Ta}$ gets sign reversed. The THz charge current can be assumed to be consisting of two contributions in the spin Hall angle, one from the bulk-like intrinsic spin-orbit coupling and another from the interface.[19] This is possible in the case of the annealed CoFeB/Ta, in which, the B-ion diffusion during the annealing process can alter the interface as compared to that in the case of the as-grown CoFeB/Ta. This possibility is schematically shown in Fig. 5(c) that there is a perfectly sharp interface in one case and an imperfect interface in the other. The transmission of the arrows through the interface here represents the spin current conversion efficiency in the two cases. The bulk-like component of the spin Hall angle in the annealed CoFeB/Ta can be assumed to be the same as that in the as-grown CoFeB/Ta. Using the temperature-dependent values of $E_{PP}$ in Figs. 2(a,b) for the as-grown CoFeB/Ta and annealed CoFeB/Ta, respectively, both of the bulk-like and the interfacial components to the $E_{PP}$ in the annealed CoFeB/Ta are shown separately in Fig. 6. Further, in the same figure, the experimentally measured values of the spin Hall angle in Ta, taken from ref.[[16]], have also been plotted for a comparison. A close match between the spin Hall angle and the bulk-like component of $E_{PP}$ in their temperature-dependent behavior, justify the initial postulate of having the bulk-like intrinsic and interfacial extrinsic components of the overall spin Hall angle/SOC in the annealed CoFeB/Ta. A giant interfacial contribution to the spin Hall angle and hence the THz amplitude $E_{PP}$ along with its sign reversal at ~200 K in the annealed CoFeB/Ta, is clearly seen in Fig. 6. A possible reason and further understanding for the observed behaviour is elaborated in the discussion part.

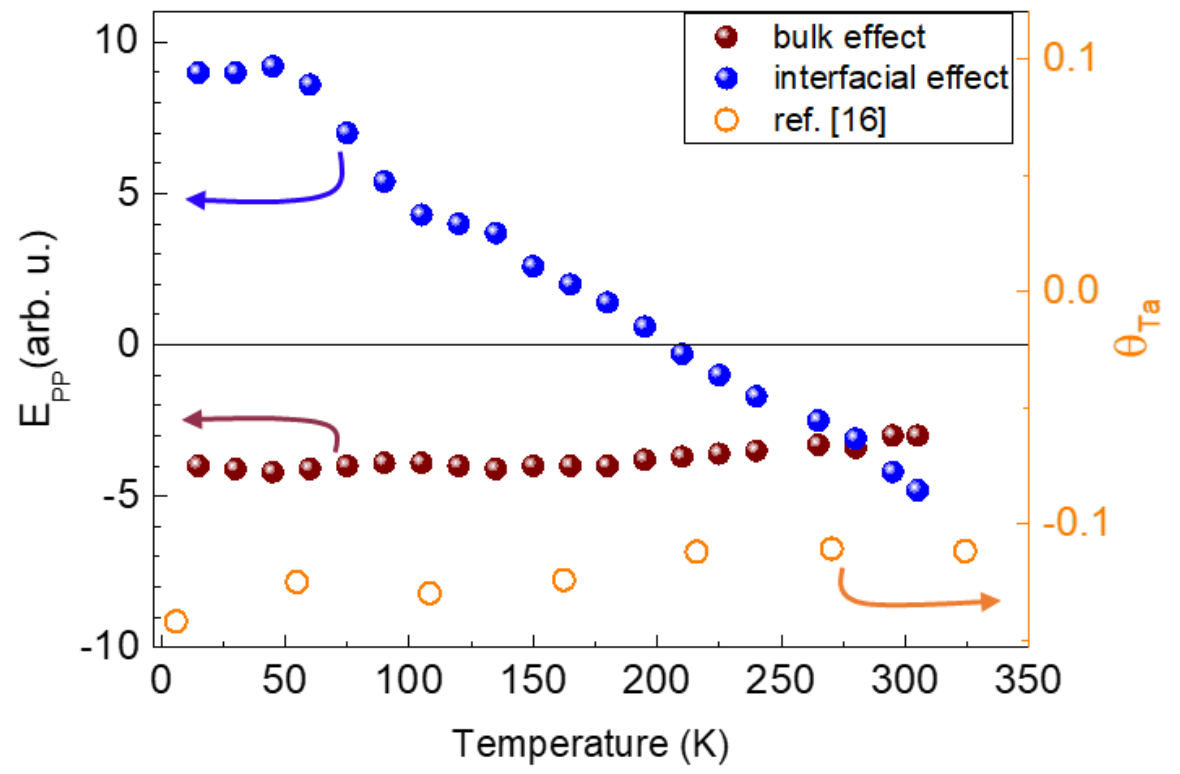

**FIGURE 6.** Temperature-dependent THz emission from the bulk-like intrinsic and interfacial extrinsic contributions of the overall SOC in the annealed CoFeB/Ta bilayer heterostructure. Data from ref. [16] for the spin Hall angle is plotted using open circles, for a comparison.

In the process of THz emission from spintronic heterostructures, both the bulk-like or the intrinsic and the interfacial or the extrinsic contributions to the SOC must be considered.[60] Generally, the ISHE is at play maximally in bulk and weakly or reasonably large, depending on the conditions at the interface between two metal layers, whereas, the IREE is majorly present in the latter case only.[28, 60] The dominance of one over the other depends on the nature of the interface. For example, in a FM/NM heterostructure using Ag/Bi layer combination for the NM, the Ag/Bi interface was well understood as a Rashba interface in which the SOC was mainly governed by the IREE.[28] However, if $Ag_xBi_{100-x}$ alloy is used as the NM layer, it was found that the interfacial contribution from ISHE can be significantly larger than the IREE for certain compositions.[27] A dominant contribution from an interface-related extrinsic mechanism over the intrinsic one for the spin-charge conversion has also been indicated in a few previous studies on other material structures. In a theoretical study[15], it was shown that the value of the spin Hall angle in pure bulk Pt can get enhanced by more than an order due to the additional contribution from its interface with a FM layer of permalloy (Py). Bismuth (Bi) has a relatively much smaller value of the intrinsic spin Hall angle than Pt; however, in a Py/Bi heterostructure, the overall spin Hall angle that is majorly contributed by the interfacial extrinsic mechanism becomes much higher than the intrinsic one in Pt alone.[19]

To probe the interface related phenomena, THz emission spectroscopy is realised to be as effective as the steady state FMR based techniques because of the interrelated total SOC and spin-orbit torque mechanisms. It is known that in presence of an interlayer, magnitude of THz emission through ISHE in a FM/NM spintronic heterostructure follows the same trend as that of the Gilbert damping parameter with respect to the thickness of the interlayer.[37] The latter was measured by using steady-state FMR technique. The spin orbit torque has two components, if disentangled, the damping-like (DL) and the field-like (FL). There can be only DL torque from the intrinsic bulk part of the SHE, while from the interface, among the DL and the FL torques, the latter dominates as seen earlier from spin-torque FMR and harmonic Hall measurements.[61-63] CoFeB/Ta is a system, where harmonic Hall measurements have revealed significant contribution from the FL torque along with the intrinsic DL contribution.[63] Therefore, the present results are indicative that the FL torque contributes differently in the as-grown and annealed CoFeB/Ta bilayers. The annealing process causes boron deficiency in CoFeB and hence the composition around the interface and its nature gets altered. The interfacial modifications in the annealed CoFeB/Ta are evident from the XRD and MH measurements also (please see Section S3 in the supplementary information). The sign reversal in the THz polarity at ~200 K arises from the interfacial effect for the extrinsic contribution to the total SOC in the annealed CoFeB/Ta (Fig. 6). This particular result would analogously correspond to sign reversal of interfacial FL torque[61, 62] in temperature-dependent spin-torque FMR and harmonic Hall measurements.

In conclusion, we have studied the spin-charge conversion process through THz emission measurements on the as-grown and annealed CoFeB/Ta bilayers spintronic THz emitters and compared them with an extensively studied Fe/Pt system. Detailed temperature-dependent measurements have helped determine the behaviour of the spin Hall resistivity and the contributions from the intrinsic and extrinsic components in it. The linear proportionality of the spin Hall resistivity with the squared resistivity of Pt and $\alpha$-phase Ta in Fe/Pt and the as-grown CoFeB/Ta, respectively, indicate the dominance of intrinsic bulk contribution to the overall spin-charge conversion mechanism in them. It has been found that the THz pulse generated from the annealed CoFeB/Ta gets polarity reversed below a certain temperature, which is solely related to the sign reversal in the interfacial spin Hall angle. This peculiar behaviour of the sign reversal of the spin Hall angle at ~200 K clearly suggests a significant role of the interfacial contributions along with the intrinsic contribution to the inverse spin Hall effect, which to our best knowledge, has not been seen hitherto. In addition, our measurements also strengthen the point further that ultrafast THz emission can be used for studying the microscopic spin-orbit coupling in the spintronic structures in a contactless and non-destructive manner through the time-domain THz spectroscopy.

## ACKNOWLEDGMENTS

SK acknowledges the Science and Engineering Research Board (SERB), Department of Science and Technology, Government of India, for financial support through project no. CRG/2020/000892. Joint Advanced Technology Center, IIT Delhi is also acknowledged for support through EMDTERA#5 project. One of the authors (Sandeep Kumar) acknowledges the University Grants Commission, Government of India for Senior Research Fellowship.

## DECLARATION

The authors declare no conflict of interest.

## REFERENCES

1. M. I. Dyakonov and V. I. Perel, Physics Letters A **35** (6), 459-460 (1971).
2. J. Sinova, S. O. Valenzuela, J. Wunderlich, C. H. Back and T. Jungwirth, Reviews of Modern Physics **87** (4), 1213-1260 (2015).
3. A. Hoffmann, IEEE Transactions on Magnetics **49** (10), 5172-5193 (2013).
4. K. Garello, I. M. Miron, C. O. Avci, F. Freimuth, Y. Mokrousov, S. Blügel, S. Auffret, O. Boulle, G. Gaudin and P. Gambardella, Nature Nanotechnology **8** (8), 587-593 (2013).
5. V. M. Edelstein, Solid State Communications **73** (3), 233-235 (1990).
6. Y. A. Bychkov and É. I. J. Z. P. R. Rashba, **39**, 66 (1984).
7. J. E. Hirsch, Physical Review Letters **83** (9), 1834-1837 (1999).
8. Y. Niimi and Y. Otani, Reports on Progress in Physics **78** (12), 124501 (2015).
9. O. M. J. van 't Erve, A. T. Hanbicki, K. M. McCreary, C. H. Li and B. T. Jonker, Applied Physics Letters **104** (17), 172402 (2014).
10. T. Kampfrath, M. Battiato, P. Maldonado, G. Eilers, J. Nötzold, S. Mährlein, V. Zbarsky, F. Freimuth, Y. Mokrousov, S. Blügel, M. Wolf, I. Radu, P. M. Oppeneer and M. Münzenberg, Nature Nanotechnology **8**, 256 (2013).
11. R. Karplus and J. M. Luttinger, Physical Review **95** (5), 1154-1160 (1954).
12. J. Smit, Physica **24** (1), 39-51 (1958).
13. L. Berger, Physical Review B **2** (11), 4559-4566 (1970).
14. E. Sagasta, Y. Omori, S. Vélez, R. Llopis, C. Tollan, A. Chuvilin, L. E. Hueso, M. Gradhand, Y. Otani and F. Casanova, Physical Review B **98** (6), 060410 (2018).
15. L. Wang, R. J. H. Wesselink, Y. Liu, Z. Yuan, K. Xia and P. J. Kelly, Physical Review Letters **116** (19), 196602 (2016).
16. Q. Hao and G. Xiao, Physical Review B **91** (22), 224413 (2015).
17. E. Sagasta, Y. Omori, M. Isasa, M. Gradhand, L. E. Hueso, Y. Niimi, Y. Otani and F. Casanova, Physical Review B **94** (6), 060412 (2016).
18. M.-H. Nguyen, D. C. Ralph and R. A. Buhrman, Physical Review Letters **116** (12), 126601 (2016).
19. D. Hou, Z. Qiu, K. Harii, Y. Kajiwara, K. Uchida, Y. Fujikawa, H. Nakayama, T. Yoshino, T. An, K. Ando, X. Jin and E. Saitoh, Applied Physics Letters **101** (4), 042403 (2012).
20. J. Cramer, T. Seifert, A. Kronenberg, F. Fuhrmann, G. Jakob, M. Jourdan, T. Kampfrath and M. Kläui, Nano Letters **18** (2), 1064-1069 (2018).
21. C. O. Avci, K. Garello, M. Gabureac, A. Ghosh, A. Fuhrer, S. F. Alvarado and P. Gambardella, Physical Review B **90** (22), 224427 (2014).
22. L. Cheng, Z. Li, D. Zhao and E. E. M. Chia, APL Materials **9** (7), 070902 (2021).
23. L. Cheng, X. Wang, W. Yang, J. Chai, M. Yang, M. Chen, Y. Wu, X. Chen, D. Chi, K. E. J. Goh, J.-X. Zhu, H. Sun, S. Wang, J. C. W. Song, M. Battiato, H. Yang and E. E. M. Chia, Nature Physics **15** (4), 347-351 (2019).
24. M. Matthiesen, D. Afanasiev, J. R. Hortensius, T. C. van Thiel, R. Medapalli, E. E. Fullerton and A. D. Caviglia, Applied Physics Letters **116** (21), 212405 (2020).
25. P. Agarwal, R. Medwal, A. Kumar, H. Asada, Y. Fukuma, R. S. Rawat, M. Battiato and R. Singh, Advanced Functional Materials **31** (17), 2010453 (2021).
26. O. Gueckstock, L. Nádvorník, M. Gradhand, T. S. Seifert, G. Bierhance, R. Rouzegar, M. Wolf, M. Vafaee, J. Cramer, M. A. Syskaki, G. Woltersdorf, I. Mertig, G. Jakob, M. Kläui and T. Kampfrath, Advanced Materials **33** (9), 2006281 (2021).
27. J. Shen, Z. Feng, P. Xu, D. Hou, Y. Gao and X. Jin, Physical Review Letters **126** (19), 197201 (2021).
28. M. B. Jungfleisch, Q. Zhang, W. Zhang, J. E. Pearson, R. D. Schaller, H. Wen and A. Hoffmann, Physical Review Letters **120** (20), 207207 (2018).
29. W. Zhang, M. B. Jungfleisch, W. Jiang, J. E. Pearson and A. Hoffmann, Journal of Applied Physics **117** (17), 17C727 (2015).
30. A. Kumar, R. Bansal, S. Chaudhary and P. K. Muduli, Physical Review B **98** (10), 104403 (2018).
31. L. Liu, C.-F. Pai, Y. Li, H. W. Tseng, D. C. Ralph and R. A. Buhrman, Science **336** (6081), 555-558 (2012).
32. S. Ikeda, K. Miura, H. Yamamoto, K. Mizunuma, H. D. Gan, M. Endo, S. Kanai, J. Hayakawa, F. Matsukura and H. Ohno, Nature Materials **9** (9), 721-724 (2010).
33. Y. Sasaki, Y. Kota, S. Iihama, K. Z. Suzuki, A. Sakuma and S. Mizukami, Physical Review B **100** (14), 140406 (2019).
34. Y. Sasaki, K. Z. Suzuki and S. Mizukami, Applied Physics Letters **111** (10), 102401 (2017).
35. T. Seifert, S. Jaiswal, U. Martens, J. Hannegan, L. Braun, P. Maldonado, F. Freimuth, A. Kronenberg, J. Henrizi, I. Radu, E. Beaurepaire, Y. Mokrousov, P. M. Oppeneer, M. Jourdan, G. Jakob, D. Turchinovich, L. M. Hayden, M. Wolf, M. Münzenberg, M. Kläui and T. Kampfrath, Nature Photonics **10**, 483 (2016).
36. H. Cheng, Y. Wang, H. He, Q. Huang and Y. Lu, Physical Review B **105** (15), 155141 (2022).

37. J. Hawecker, T.-H. Dang, E. Rongione, J. Boust, S. Collin, J.-M. George, H.-J. Drouhin, Y. Laplace, R. Grasset, J. Dong, J. Mangeney, J. Tignon, H. Jaffrès, L. Perfetti and S. Dhillon, Advanced Optical Materials **9** (17), 2100412 (2021).
38. S. Kumar, A. Nivedan, A. Singh, Y. Kumar, P. Malhotra, M. Tondusson, E. Freysz and S. Kumar, iScience **24** (10), 103152 (2021).
39. J. Degert, M. Tondusson, V. Freysz, E. Abraham, S. Kumar and E. Freysz, Opt. Express **30** (11), 18995-19004 (2022).
40. S. Kumar, A. Singh, S. Kumar, A. Nivedan, M. Tondusson, J. Degert, J. Oberlé, S. J. Yun, Y. H. Lee and E. Freysz, Opt. Express **29** (3), 4181-4190 (2021).
41. S. Kumar, A. Singh, A. Nivedan, S. Kumar, S. J. Yun, Y. H. Lee, M. Tondusson, J. Degert, J. Oberle and E. Freysz, Nano Select **2** (10), 2019-2028 (2021).
42. S. Kumar, A. Nivedan, A. Singh and S. Kumar, Pramana **95** (2), 75 (2021).
43. X. Xin, H. Altan, A. Saint, D. Matten and R. R. Alfano, Journal of Applied Physics **100** (9), 094905 (2006).
44. S. Kumar and S. Kumar, Applied Physics Letters **120** (20), 202403 (2022).
45. L. Scheuer, M. Ruhwedel, D. Karfaridis, I. G. Vasileiadis, D. Sokoluk, G. Torosyan, G. Vourlias, G. P. Dimitrakopoulos, M. Rahm, B. Hillebrands, T. Kehagias, R. Beigang and E. T. Papaioannou, iScience **25** (5), 104319 (2022).
46. K. P. H. Lui and F. A. Hegmann, Journal of Applied Physics **93** (11), 9012-9018 (2003).
47. G. Torosyan, S. Keller, L. Scheuer, R. Beigang and E. T. Papaioannou, Scientific Reports **8** (1), 1311 (2018).
48. E. T. Papaioannou and R. Beigang, Nanophotonics **10** (4), 1243-1257 (2021).
49. M. Cecot, Ł. Karwacki, W. Skowroński, J. Kanak, J. Wrona, A. Żywczak, L. Yao, S. van Dijken, J. Barnaś and T. Stobiecki, Scientific Reports **7** (1), 968 (2017).
50. H. Gamou, Y. Du, M. Kohda and J. Nitta, Physical Review B **99** (18), 184408 (2019).
51. Y.-Y. Chen and J.-Y. Juang, Measurement Science and Technology **27** (7), 074006 (2016).
52. C. O. Avci, K. Garello, C. Nistor, S. Godey, B. Ballesteros, A. Mugarza, A. Barla, M. Valvidares, E. Pellegrin, A. Ghosh, I. M. Miron, O. Boulle, S. Auffret, G. Gaudin and P. Gambardella, Physical Review B **89** (21), 214419 (2014).
53. T. S. Seifert, N. M. Tran, O. Gueckstock, S. M. Rouzegar, L. Nadvornik, S. Jaiswal, G. Jakob, V. V. Temnov, M. Münzenberg, M. Wolf, M. Kläui and T. Kampfrath, Journal of Physics D: Applied Physics **51** (36), 364003 (2018).
54. P. Mohn and E. P. Wohlfarth, Journal of Physics F: Metal Physics **17** (12), 2421-2430 (1987).
55. Y. Tian, L. Ye and X. Jin, Physical Review Letters **103** (8), 087206 (2009).
56. M. Isasa, E. Villamor, L. E. Hueso, M. Gradhand and F. Casanova, Physical Review B **92** (1), 019905 (2015).
57. M. Isasa, E. Villamor, L. E. Hueso, M. Gradhand and F. Casanova, Physical Review B **91** (2), 024402 (2015).
58. T. Tanaka, H. Kontani, M. Naito, T. Naito, D. S. Hirashima, K. Yamada and J. Inoue, Physical Review B **77** (16), 165117 (2008).
59. D. Qu, S. Y. Huang, G. Y. Guo and C. L. Chien, Physical Review B **97** (2), 024402 (2018).
60. C. Zhou, Y. P. Liu, Z. Wang, S. J. Ma, M. W. Jia, R. Q. Wu, L. Zhou, W. Zhang, M. K. Liu, Y. Z. Wu and J. Qi, Physical Review Letters **121** (8), 086801 (2018).
61. Y. Du, H. Gamou, S. Takahashi, S. Karube, M. Kohda and J. Nitta, Physical Review Applied **13** (5), 054014 (2020).
62. C.-F. Pai, Y. Ou, L. H. Vilela-Leão, D. C. Ralph and R. A. Buhrman, Physical Review B **92** (6), 064426 (2015).
63. G. Allen, S. Manipatruni, D. E. Nikonov, M. Doczy and I. A. Young, Physical Review B **91** (14), 144412 (2015).

# Supporting Information

### S.1 Strength of the emitted THz pulse electric field

The THz electric field emitted from the spintronic emitters in the experiments can be reconstructed from the measured electro-optic signal sampled on a ZnTe crystal by following the standard method.[1-3] According to this, the THz electric field can be expressed in terms of the differential intensity measured on the balanced photodiode as following

$$E_{THz}\left(\frac{V}{cm}\right) = \left(\frac{\Delta I}{I}\right)\frac{2c}{\omega n^3 r_{41} L}(\cos\alpha\ \sin 2\beta + 2\sin\alpha\ \cos 2\beta) \tag{S1}$$

Here, c = 3x10$^8$ m/s is the speed of light in vacuum, ω = 2πν = 2πc/800nm is the angular frequency of the gating beam as shown in Fig. 1 of the main paper. The thickness of the ZnTe crystal was L = 0.5mm. The optical constants of the crystal are: refractive index,[3] n = 2.85 and nonlinear coefficient,[1] $r_{41}$ = 3.9x10$^{-12}$ m/V. α and β are the angles of THz beam polarization and gating beam polarization with respect to the c-axis (001) of the ZnTe crystal, which are kept at 90$^0$ and 180$^0$, respectively, in our experiment. With all these values, Eq. S1 reduces to $E^{THz} \sim 28\left(\frac{\Delta I}{I}\right)\left(\frac{kV}{cm}\right)$. The quantity $\Delta I/I$ represents the differential change in the gating beam intensity on the balanced photo detector due to the presence of the THz beam. The above calculation is for the THz electric field at the detector position. For the exact determination of the emitted THz electric field just after the emitter, it requires to include the details of the transmission and loss factors of various optical components[2] (e.g. 3000 ohm.cm HR-Si wafer, polypropylene cryostat window, 15 cm focal length parabolic mirrors, ZnTe detector) used in the THz beam path. By appropriately considering all the above factors, an estimation of the THz electric field amplitude emitted from the spintronic emitter has been made. The magnitude of the spectrally integrated THz signal emitted from the efficient spintronic emitters is quite comparable with that from a standard nonlinear electro-optic crystal[4], which can be determined by simply interchanging the two emitters in the same experimental setup. These have even served as a standard for comparing the THz generation efficiency of other material structures.[2]

### S.2 Intrinsic Spin-Hall conductivity from the emitted THz electric field strength

According to Eq. 2 of the main manuscript, the spin Hall resistivity in terms of THz electric field can be written as

$$\rho_{SH} = E_{THz}\left(\frac{1}{d}\cdot\frac{\rho_{FM/NM}}{\rho_{NM}}\right)^{-1}.\ \frac{1}{J_s} \tag{S2}$$

Also, the spin Hall resistivity has a linear relation with the squared longitudinal resistivity, ($\rho_{NM}^2$), whose slope is equal to the intrinsic spin Hall conductivity of the NM layer (see Eq. 3 of main text). Hence, intrinsic spin Hall conductivity can be rewritten in the form of the THz electric field as follows

$$\sigma_{NM}^{int.}.\rho_{NM}^2 = E_{THz}\left(\frac{1}{d}\cdot\frac{\rho_{FM/NM}}{\rho_{NM}}\right)^{-1}.\ \frac{1}{J_s} \tag{S3}$$

or

$$\sigma_{NM}^{int.} = \frac{E_{THz}\left(\frac{1}{d}\frac{\rho_{FM/NM}}{\rho_{NM}}\right)^{-1}}{\rho_{NM}^2}.\ \frac{1}{J_s} \tag{S4}$$

where, the first term in Eq. S4 is nothing but the value of the *slope* obtained from the $E_{THz}\left(\frac{1}{d}\cdot\frac{\rho_{FM/NM}}{\rho_{NM}}\right)^{-1}$ versus $\rho_{NM}^2$ plot as discussed in the main manuscript. $J_s$ is the spin current density generated from the ultrafast optical excitation of the FM/NM spintronic heterostructure. Assuming superdiffusive process, the spin current density averaged over the thickness ($J_s/d$) in such heterostructures[5] is typically ~10$^{30}$ A/m$^2$. By considering the above, Eq. S4 can be rewritten as

$$\sigma_{NM}^{int.} = (slope).\ \frac{10^2}{6.58}\left(\frac{\hbar}{e}\right)(\Omega^{-1}cm^{-1}) \tag{S5}$$

Equation S5 has been used to calculate the intrinsic spin Hall conductivity for all our samples. The exact value of the intrinsic spin Hall conductivity may differ because we have used the theoretically provided value of the spin current density and also the underestimated temperature variation of spin relaxation length and spin current density.

### S.3 Structural and magnetic properties of the as-grown and annealed CoFeB/Ta systems

The X-ray diffraction (XRD) measurements were carried out using a PANalytical X'Pert diffractometer with a Cu-$K_\alpha$ source on CoFeB/Ta thin films. Figure S3 (a) shows the GIXRD (grazing incident angle X-ray diffraction) patterns of the as-grown and annealed CoFeB/Ta heterostructure sample grown on the silicon substrate. The diffraction peaks observed in Fig. S3(a) at angles of ~$38^0$ and ~$55^0$ in the case of an as-grown sample are due to the $\alpha$-phase Ta and the substrate. The absence of any other peak indicates the amorphous phase of CoFeB in the as-grown CoFeB/Ta heterostructure. For the annealed CoFeB/Ta sample, an additional Bragg peak at ~$45^0$ angle is observed, which is related to the bcc crystal structure of CoFe, evidently suggesting that boron diffusion at the interface with the adjacent Ta layer has taken place.[6] The above fact is consistently seen in the magnetic hysteresis (M-H) measurements performed on both the as-grown and annealed CoFeB/Ta samples. The substrate-corrected hysteresis loops are shown in Fig. S3(b). The hysteresis curve of the annealed sample shows a larger value of both the coercive field and the saturation magnetization than the as-grown CoFeB/Ta sample. This observation is attributed to the recrystallization of the CoFeB layer in the annealed CoFeB/Ta heterostructure due to the process of annealing at ~625 K in our experiments. The observed MH-behaviour of the as-grown and the annealed CoFeB/Ta samples is consistent with the literature.[7, 8]

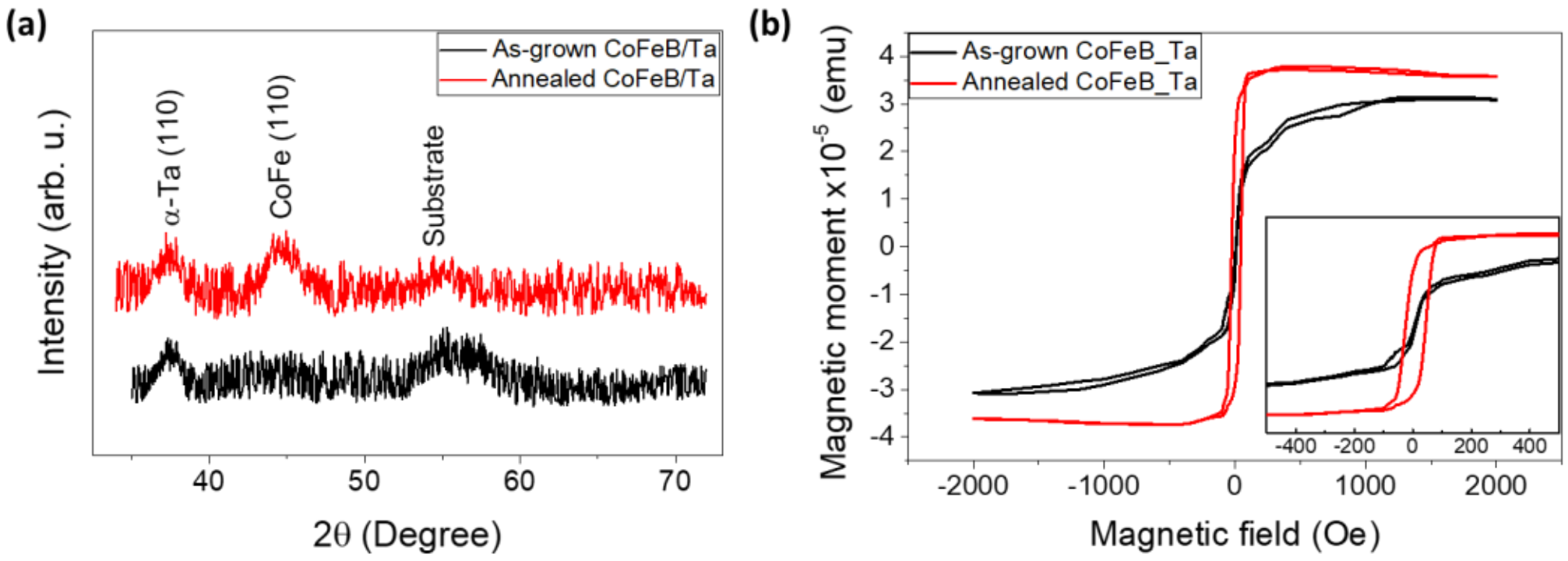

Figure S3. (a) XRD plots of as-grown and annealed CoFeB/Ta spintronics heterostructures. The peaks and corresponding crystalline planes are indicated. (b) In-plane hysteresis loops from MH measurement on the as-grown and annealed CoFeB/Ta heterostructures. Inset: zoomed-in view for clarity.

### References

1. P. C. M. Planken, H.-K. Nienhuys, H. J. Bakker and T. Wenckebach, J. Opt. Soc. Am. B **18** (3), 313-317 (2001).
2. L. Cheng, X. Wang, W. Yang, J. Chai, M. Yang, M. Chen, Y. Wu, X. Chen, D. Chi, K. E. J. Goh, J.-X. Zhu, H. Sun, S. Wang, J. C. W. Song, M. Battiato, H. Yang and E. E. M. Chia, Nature Physics **15** (4), 347-351 (2019).
3. D. Yang, J. Liang, C. Zhou, L. Sun, R. Zheng, S. Luo, Y. Wu and J. Qi, Advanced Optical Materials **4** (12), 1944-1949 (2016).
4. H. H. Li, Journal of Physical and Chemical Reference Data **13** (1), 103-150 (1984).
5. T. Kampfrath, M. Battiato, P. Maldonado, G. Eilers, J. Nötzold, S. Mährlein, V. Zbarsky, F. Freimuth, Y. Mokrousov, S. Blügel, M. Wolf, I. Radu, P. M. Oppeneer and M. Münzenberg, Nature Nanotechnology **8**, 256 (2013).
6. H. Bouchikhaoui, P. Stender, D. Akemeier, D. Baither, K. Hono, A. Hütten and G. Schmitz, Applied Physics Letters **103** (14), 142412 (2013).
7. C. Burrowes, N. Vernier, J. P. Adam, L. Herrera Diez, K. Garcia, I. Barisic, G. Agnus, S. Eimer, J.-V. Kim, T. Devolder, A. Lamperti, R. Mantovan, B. Ockert, E. E. Fullerton and D. Ravelosona, Applied Physics Letters **103** (18), 182401 (2013).
8. G. V. Swamy, H. Pandey, A. K. Srivastava, M. K. Dalai, K. K. Maurya, Rashmi and R. K. Rakshit, AIP Advances **3** (7), 072129 (2013).